\documentclass[preprint]{aastex} 
\def\etal{{\it et al.}}
\def\lae{\mathrel{<\kern-1.0em\lower0.9ex\hbox{$\sim$}}}
\def\gae{\mathrel{>\kern-1.0em\lower0.9ex\hbox{$\sim$}}}
\def\deg{^{\circ}}

\def\mone{$^{-1}$}

\newcommand{\Msun}{$M_{\sun}$}
\newcommand{\Zsun}{$Z_{\sun}$}

\begin{document}

\title{ 3C~236: Radio Source, Interrupted? }

\author{Christopher P. O'Dea, Anton M. Koekemoer, Stefi A. Baum, William 
B. Sparks, Andre\'\  R. Martel, Mark G. Allen, 
\& Ferdinando D. Macchetto}
\affil{ Space Telescope Science Institute\footnote{Operated
by the Association of Universities for Research in
Astronomy, Inc. under contract NAS 5-26555 with the National Aeronautics and
Space Administration.}
}
\authoraddr{ 3700 San Martin Dr., Baltimore, MD 21218 }
\author{George K. Miley }
\affil{Sterrewacht Leiden}
\author{Astronomical Journal, in press} 

\begin{abstract}
We present new HST STIS/MAMA near-UV  images and archival WFPC2 
V and R band images which reveal the presence of four star forming 
regions in an arc along the edge of the  dust lane in the giant 
(4 Mpc) radio galaxy 3C~236. 
Two of the star forming regions are relatively young with ages of 
order $\sim 10^7$ yr, while the other two are older with ages of order 
$\sim 10^8 -  10^9$ yr which is comparable to the estimated age 
of the giant radio source. 

Based on dynamical and spectral aging arguments, we suggest that 
the fuel supply to the AGN was interrupted for $\sim 10^7$ yr and has
now been restored, resulting in the formation of the inner
2 kpc scale radio source. This time scale is similar to that of 
the age of the youngest of the star forming regions.

We suggest that the transport of gas in the disk is non-steady and that
this produces both the multiple episodes of star formation in the disk
as well as the multiple epochs of radio source activity. If the
inner radio source and the youngest star forming region are related
by the same event of gas transport, the gas must be transported
from the hundreds of pc scale to the sub-parsec scale on a time scale
of $\sim 10^7$ yr, which is similar to the dynamical time scale of
the gas on the hundreds of pc scales.

\end{abstract}


\section{INTRODUCTION}

3C~236 is a powerful  classical double radio galaxy  
at relatively low redshift (z=0.1) (Willis, Strom, \& Wilson 1974;
Barthel \etal\ 1985; Schilizzi \etal\ 2000). Its angular size of 40 arcmin
corresponds to a projected linear size of 4 Mpc (possibly 4.5 Mpc 
deprojected -- Schilizzi \etal\ 2000) making it the largest
known radio galaxy and even one of the largest known objects
in the universe.\footnote{We adopt a Hubble constant of
$H_o = 75$ km s\mone\ Mpc\mone\ and
a deceleration parameter of $q_o = 0.1$ and at the redshift of
3C~236 ($z=0.10050$) the scale is 1.68 kpc arcsec\mone .}
There are no large scale radio jets detected in this source. The bright
lobes do not connect to the core; however at low frequencies
faint and diffuse bridges are seen (Mack \etal\ 1997;  Waldram \&
Riley 1993). The nuclear radio source appears to be a small 
classical double source with an extent of 2 kpc and thus resembles
a Compact Steep Spectrum (CSS)\footnote{See O'Dea (1998) for a review 
of Compact Steep Spectrum and GHz Peaked Spectrum radio sources.} 
radio source (Fomalont, Miley, \& Bridle 1979; Barthel \etal\ 1985). 
Van Gorkom \etal\ (1989) detected the 21 cm line of HI in absorption against 
the nuclear radio source (This has since been imaged with VLBI by
Conway 1999).
One of the most remarkable features of 3C 236 is the alignment of the 
inner 2 kpc component and with the overall 4 Mpc source structure to within 
a few degrees (Fomalont and Miley 1975), 
implying that the nuclear activity axis remains constant over times 
considerably in excess of $10^7$ yr. The observed constancy in direction was 
one of the earliest indications that rotating black holes power 
extragalactic radio sources.

The presence of the 2 kpc radio source inside the 4 Mpc radio source
suggests that 3C~236 is a member of the class of ``double-double" 
radio sources (see, e.g., Schoenmakers \etal\ 1999, 2000a,b; Kaiser, 
Schoenmakers \& R\"ottgering 2000; Lara \etal\ 1999). Baum \etal\ (1990) have  
suggested 
that these sources are produced when (1) the radio activity is renewed after
a period of dormancy so that the smaller and younger source is propagating
outwards amidst the relic of the previous epoch of activity and/or (2)
the radio activity is temporarily interrupted - perhaps by ``smothering"
of the radio source by infall of gas.\footnote{Christiansen (1973) has
discussed repetitive activity in the context of plasma cloud models
and Bridle, Perley \& Henriksen (1986) have discussed a restarting jet
model for 3C219.}
In the latter case the nuclear jets
must propagate outwards though the infalling dense gas and reestablish
a connection to the lobes. 
Hooda, Mangalam, \& Wiita (1994) have presented numerical simulations which
show that jets with low Mach numbers ($M<3$) may ``stall" due to
an instability (even if the jet is still being continuously  fed by 
the nucleus). The stalled jet creates a new hot spot within the 
lobe of the source which then continues to propagate outwards.

Martel \etal\ (1999) presented an HST/WFPC2 image in the broad red (F702W)
filter of the host galaxy of 3C~236. They show that the galaxy contains
a broad dust lane and circumnuclear knots of emission. The properties
of the dust are discussed further by De Koff \etal\ (2000).  
In this paper
we present HST/WFPC2 V band (F555W) and STIS/NUV-MAMA near UV images
which show that these knots are very blue and are likely to be young
star clusters. We discuss the implications of the star formation for 
the fueling and life cycle of the radio activity.

\section{HST WFPC2 and STIS OBSERVATIONS}

\begin{figure}
\plotone{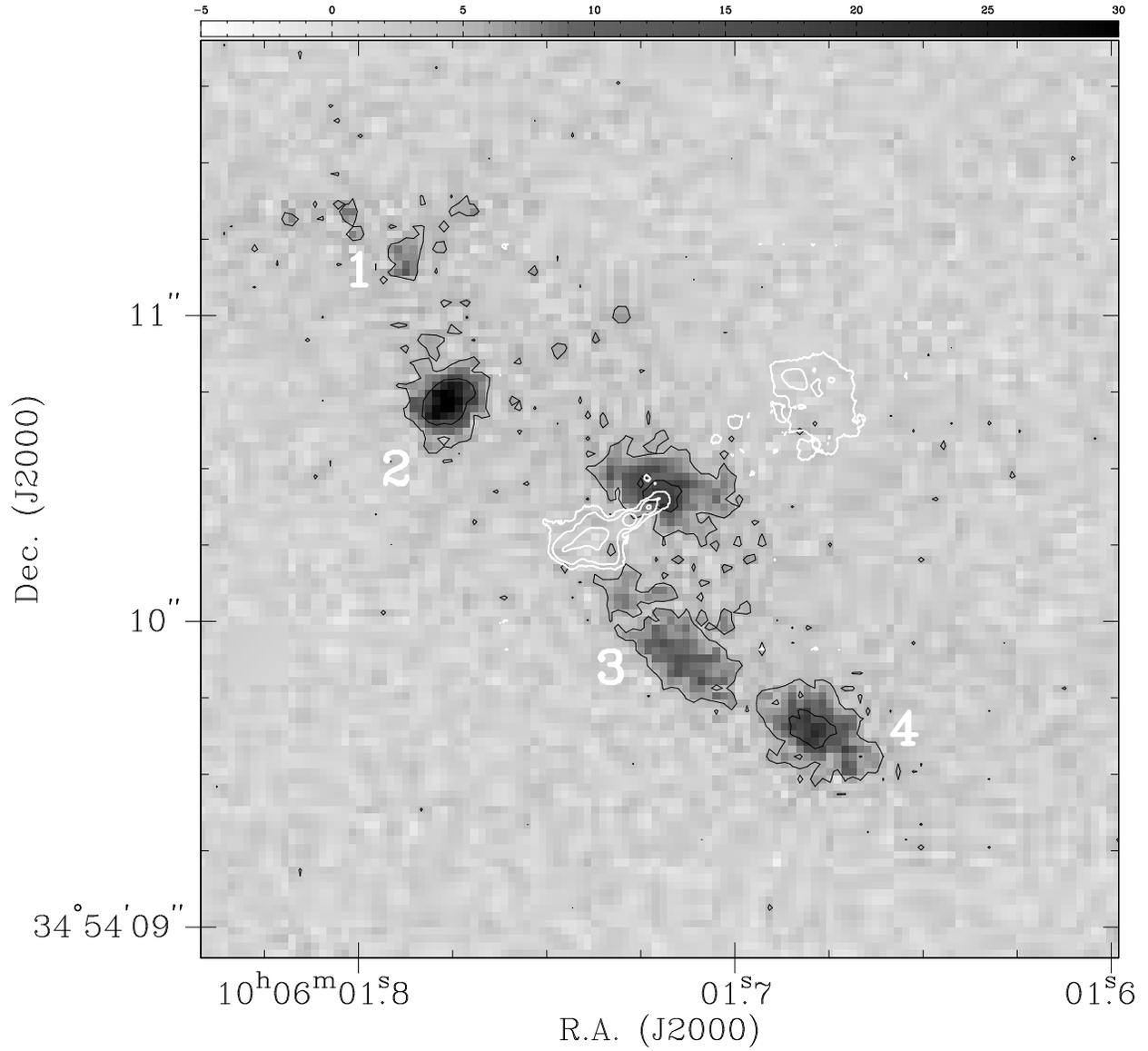}
\caption{3C~236. Overlay of radio source contours (global VLBI 1.66 GHz image 
from Schilizzi \etal\ 2000) on our STIS near-UV  image (grey scale). The four
near-UV extra-nuclear features are labeled. }
\label{figmamajetoverlay}
\end{figure}

\begin{figure}
\epsscale{0.8}
\plotone{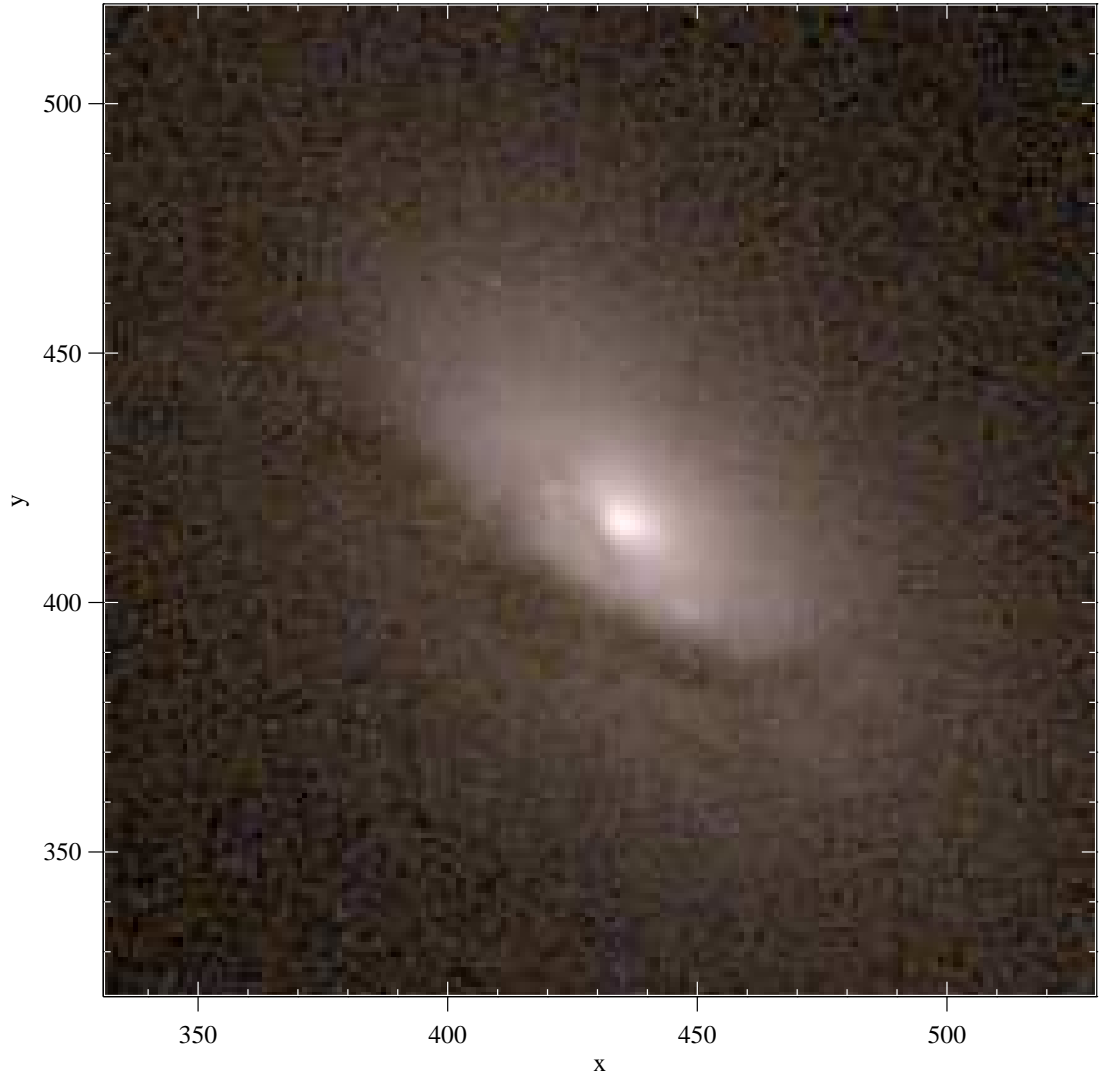}
\caption{Three-color image of 3C~236, created by assigning blue to the
STIS/NUV-MAMA image, green to the WFPC2/F555W image, and red to the WFPC2/F702W
image. Note in particular the presence of the extended dust lane to the south
and south-east of the nucleus, as well as the more compact ``disk'' of
obscuration closer to the nucleus at a position angle of $\sim 30\arcdeg$.}
\label{figcolor}
\end{figure}

The HST observations were obtained as part of a snapshot survey of
the 3CR sample by Sparks and collaborators (e.g., De Koff \etal\ 1996;
McCarthy \etal\ 1997; Martel \etal\ 1999). 3C236 was observed on
the PC of WFPC2  (Trauger \etal\ 1994). The 
F555W and F702W data were reduced as described by Martel  
\etal\ (1999).  The STScI calibration pipeline was re-run using updated 
reference files. Cosmic rays were removed by combining both images with the
CRREJ task in the STSDAS package in IRAF. 
The STIS NUV-MAMA (Kimble \etal\ 1998) images were reduced as described by 
Allen \etal\ (2000). Again the images were run though the STScI calibration 
pipeline using the best available reference files. The parameters of the 
observations are given in Table 1. Our STIS NUV-MAMA image is presented in
Figure~\ref{figmamajetoverlay} and a three color image created by combining 
the STIS and WFPC2 data is shown in Figure~\ref{figcolor}.

\section{RESULTS}


Our main new observational result is the detection of several very blue
resolved knots on the eastern edge of the dust lane (Figures 
\ref{figmamajetoverlay} 
and \ref{figknots}). First we set the stage by discussing the  
large-scale dust lane that extends across much of the galaxy along its 
major axis; and the inner ``disk'' of dust. 
We then discuss the properties of the very 
blue knots that we have detected with STIS, including their colors,
luminosities and inferred star-formation characteristics; 
and finally the relationship between the radio continuum emission and the
optical/UV morphology of the galaxy.

\subsection{Large-Scale Dust Lane and Inner Dust Disk}

\begin{figure}
\plotone{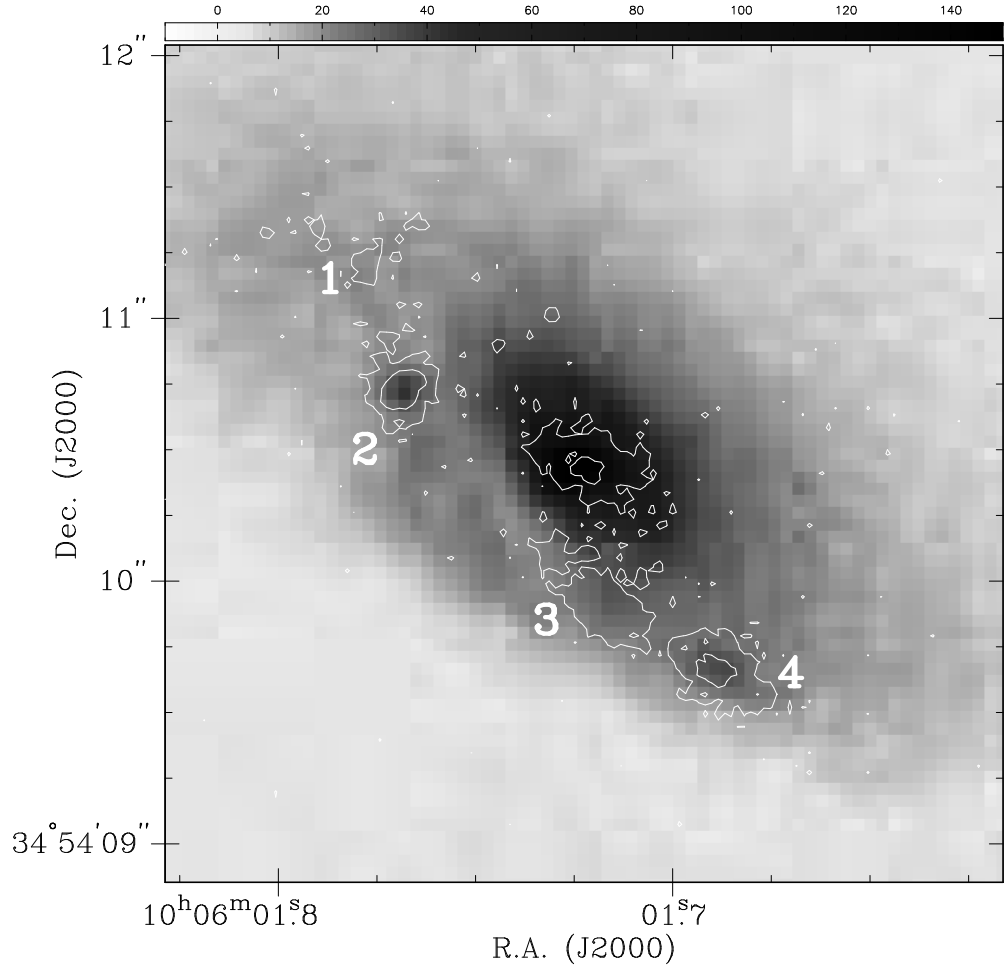}
\caption{3C~236. Overlay of STIS near-UV image (contours) on WFPC2 F555W 
V-band image (grey scale).  }
\label{figknots}
\end{figure}

\begin{figure}
\plotone{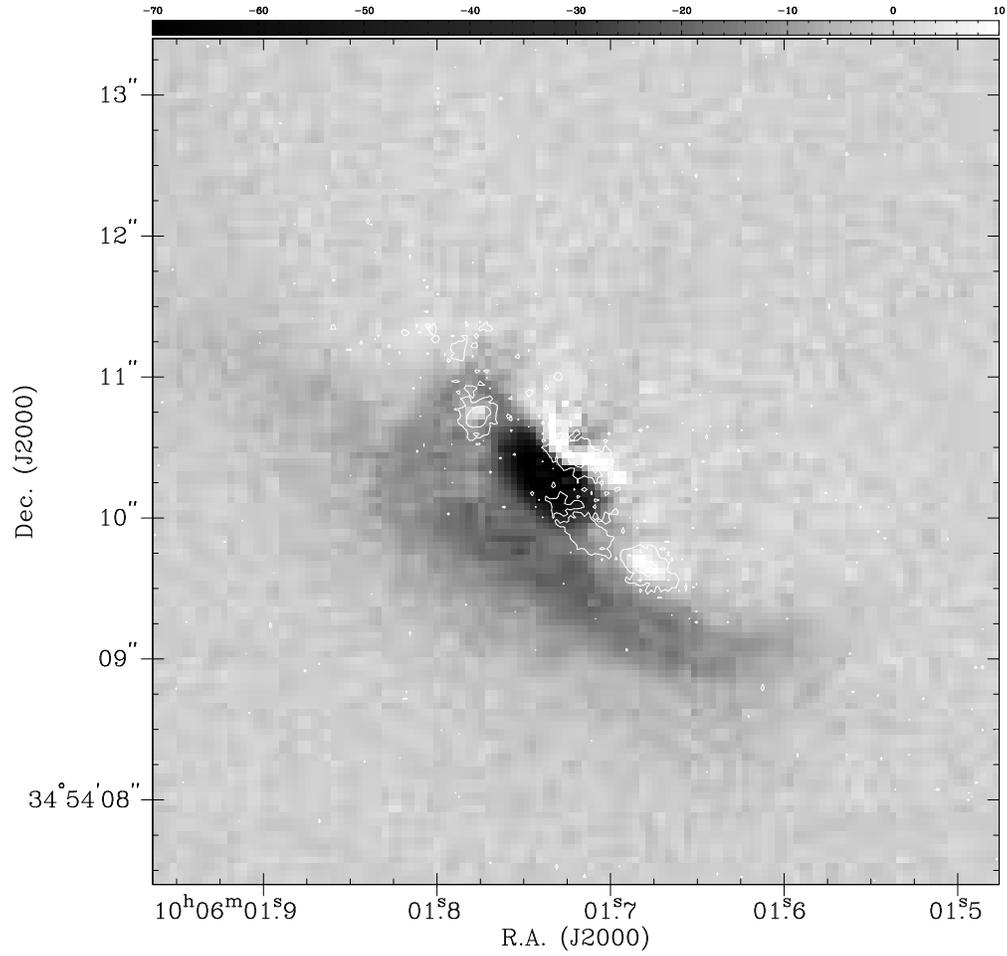}
\caption{3C~236. WFPC2 F702W (R band) image with a fit to the elliptical 
isophotes
subtracted to show the absorption due to the dust lane. Contours of the
STIS near-UV image are superposed to show the location of the near-UV knots.
 }
\label{figdustlane}
\end{figure}

In the broad-band images (e.g., Martel \etal\ 1999) there are significant 
signs of asymmetric obscuration by a system of dust lanes (with an 
overall PA$\simeq 55\deg$) across the galaxy. 
Figures \ref{figknots} and \ref{figdustlane} show the relationship of 
the blue knots to the dust lane. In particular, the F555W image
(Figure~\ref{figknots}) clearly shows a deficit of galaxy continuum 
immediately south of the blue knots, relative to the continuum level at 
corresponding radial distances in the north. Asymmetric deficits in the 
underlying elliptical morphology are also evident to the east and 
north-east of the galaxy nucleus.

\begin{figure}
\plotone{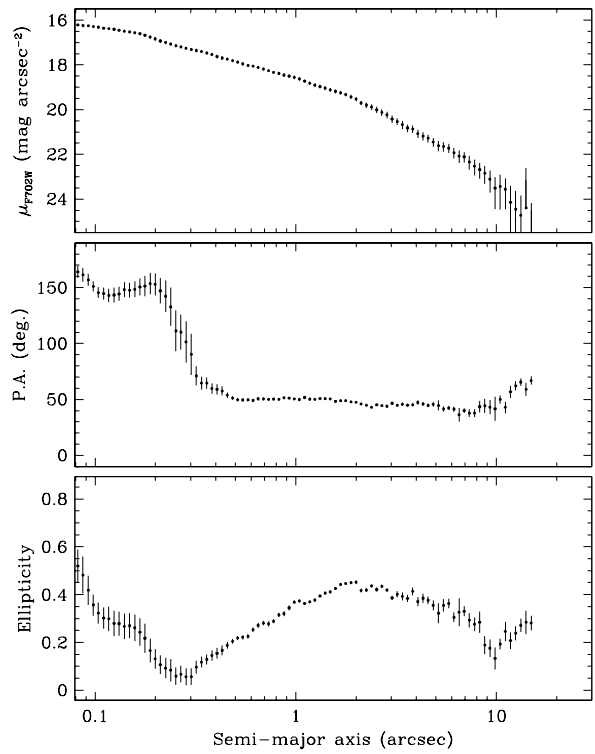} 
\caption{3C~236. The results of the the fit of elliptical isophotes 
to the WFPC2 F702W (R band) image.  We show the model parameters as 
a function of semi-major axis. (Top) Surface
brightness profile. (Center) Position angle. (Bottom) Ellipticity. 
 }
\label{figellipse}
\end{figure}

To investigate more quantitatively the extent and depth of the obscuration,
we used the IRAF ``ellipse'' task to fit a set of elliptical isophotes to the
underlying continuum in the F702W image. This band should be less susceptible
to obscuration than the F555W image and thereby better constrain the fit. We
also imposed additional constraints on the fit, including fixing the center of
the isophotes (thus their symmetry about the nucleus) and masking out obscured
regions in order to minimize their impact on the fit. The parameters of
the ellipse fit are shown in Figure~\ref{figellipse}.  

\begin{figure}
\plotone{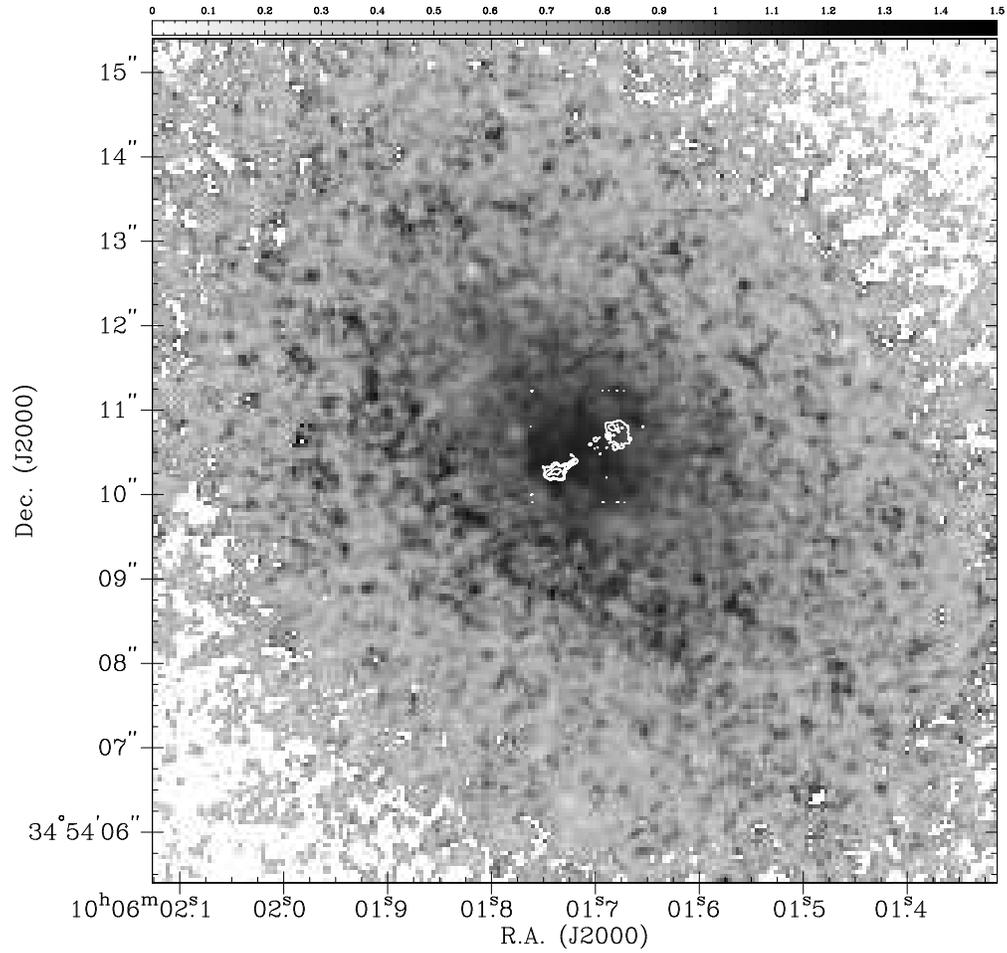}
\caption{3C~236. Left. Overlay of radio image contours (global VLBI 1.66 GHz
image from Schilizzi \etal\
2000) on the WFPC2 F555W-F702W image (grey scale). The colors are calibrated
in the HST VEGAMAG system. }
\label{fig555w702w}
\end{figure}

In Figure~\ref{figdustlane} we show the residual image produced after fitting
the elliptical model to the underlying R-band galaxy continuum emission and
subtracting this fit from the image. A large band of obscuration $\sim 10$~kpc
in extent is clearly visible along the major axis of the galaxy. This can also
be seen in the F555W-F702W color image of the galaxy (Fig~\ref{fig555w702w}).
The residual image also indicates substantial obscuration relatively close to
the nucleus, along with additional arcs and loops of obscuration to the east
and north-east of the nucleus. The implications of these features are discussed
further in Section~\ref{discussion}.

Our color image (Figure~\ref{figcolor}) shows what appears to be
an inner narrow dust disk in projection at a position angle of
$PA \simeq 30\deg$. (A hint of this is also seen in the absorption
image of De Koff \etal\ 2000). The disk has a projected size of
 $\simeq$0\farcs6 (1 kpc). This inner disk is about 5 degrees from perpendicular 
to the inner radio source ($PA \simeq 115\deg$).

\subsection{Properties of the near-UV Knots}

We detect a number of extended regions in our STIS/NUV-MAMA image
(Figures \ref{figknots} and \ref{figdustlane}). There are four very 
blue regions 
located in an arc along the eastern edge of the dust lane
at a distance of $\sim 0\farcs5$ (800 pc) from the nucleus. 
The regions are resolved with sizes $\sim 0\farcs3$ (500 pc). The properties
of the knots are given in  Table~\ref{tab:fluxes} (calibrated in the HST 
VEGAMAG system - where the magnitude of Vega in the WFPC2 filters is
defined to be zero). At the redshift of 3C236 (z=0.1005)
there are no bright emission lines present in our F25SRF2 filter. 
The knots are located on the edge of the dust lane and are not related
to the radio source or to an ionization cone. The most sensible
explanation for the knots is that they are regions of recent
and/or current star formation.

\begin{figure}
\epsscale{0.8}
\plotone{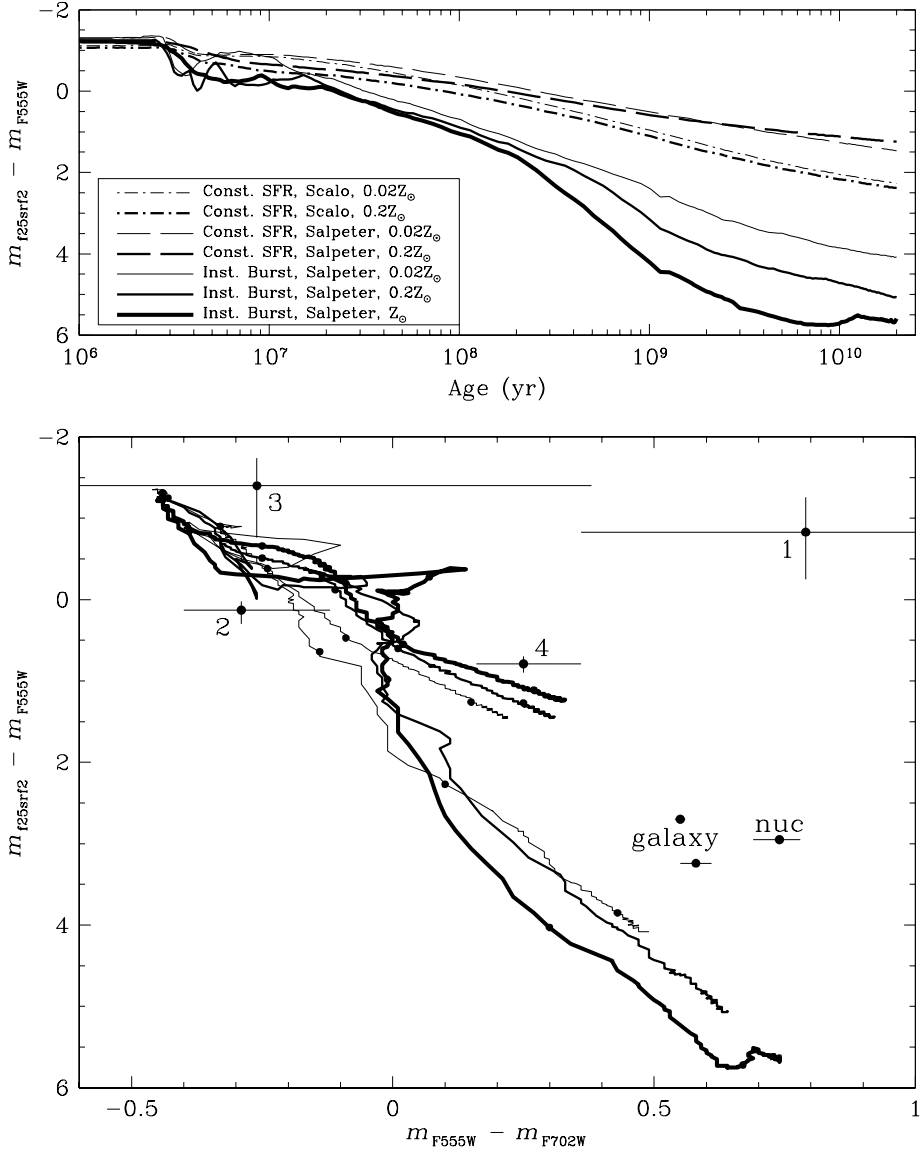}
\caption{Blue knots in 3C~236. (Top). The UV-V color (in VEGAMAGs) as a 
function of time for Bruzual and Charlot stellar population synthesis 
models for  a range of parameters. 
We show evolution of models with both constant star formation as well as an 
instantaneous burst, for metallicities of \Zsun, 0.2\Zsun\ and 0.02\Zsun, 
and for Scalo and Salpeter IMFs.  
(Bottom). The evolution in color-magnitude space of three population 
synthesis models
with a Salpeter IMF, metallicities of \Zsun, 0.2\Zsun\ and 0.02\Zsun,
together with the observed colors of the blue
clusters, as well as the colors of the surrounding galaxy and the nuclear
region. 
 }
\label{figcharlot.ps}
\end{figure}

We compare the measured properties of the blue continuum to a set of stellar
population synthesis models (Bruzual \& Charlot 1993; Charlot \& Bruzual 2000),
which have been K-corrected to the redshift of 3C~236 and calibrated in the
HST VEGAMAG system using our HST filter bandpass transmission curves.
We find it preferable to retain our measured magnitudes in the VEGAMAG system
and compare them with models that have also been calculated in this system,
rather than attempting to transform the magnitudes to other systems such as the
Johnson/Cousins system, because uncertainties in the color terms can translate
into errors of several tenths of a magnitude. The models are shown in
Figure~\ref{figcharlot.ps} and represent two scenarios for star formation:
(1)~a constant star formation rate (SFR), assuming an invariant IMF; (2)~an
instantaneous burst of star formation (or ``Simple Stellar Population'', SSP).
In each case we investigated two standard IMFs for which models were available,
namely Salpeter (1955)	
and Scalo (1986),		
each with upper and lower mass cutoffs of 0.1 and 100~\Msun\ respectively.
We also examined models of different metallicities, specifically 
Abundance $Z = $\Zsun ,
0.2\Zsun , and 0.02\Zsun\ (where \Zsun\ is the solar abundance). 
Further details of the model parameterizations are
described in Charlot \& Bruzual (1991, 2000)	
and Bruzual \& Charlot (1993).	
Having calculated HST colors and magnitudes for these various models, we
note that the difference between the color evolution of \Zsun\ and 0.2\Zsun\
models is minimal for the constant SFR models, thus for the purposes of clarity
we only plot the 0.2\Zsun\ models. For the SSP models, the color evolution is
more sensitive to metallicity but is relatively insensitive to the choice of
IMF, therefore the SSP models that we plot are for the Salpeter IMF only.

In Figure~\ref{figcharlot.ps} we plot the evolution in color-age and 
color-color space of three population synthesis models
with a Salpeter IMF, metallicities of \Zsun, 0.2\Zsun\ and 0.02\Zsun,
together with the observed blue
clusters, as well as the surrounding galaxy and the nuclear
region. The comparison suggests that the colors of the two bluest clusters
are consistent with ages $\lesssim 5 - 10$~Myr and metallicities above
$\sim 0.2Z_{\sun}$, with ages of the two redder of these clusters ranging
up to a maximum of $\sim 100$~Myr.
The reddest of the four clusters is either much older ($\sim 10^9$~yr) with a
mass $\sim 10^6$~\Msun, or otherwise is the same age as the other three, but
has higher extinction and/or a metallicity significantly above solar (however,
this cluster is fainter than the other three and its properties are not as well
constrained). In \S~\ref{discussion} we discuss in further detail the
implications of the measured properties of the blue continuum regions for
various physical mechanisms that can be responsible for triggering the star
formation.

In Table~\ref{tab:starform} we present the inferred star formation
parameters that are required to match the observed range of colors and
integrated absolute magnitude of the extended blue continuum distribution,
for the various star formation scenarios that we have described. For each
model, we tabulate the upper and lower limits of the range of epochs in its
evolution that reproduce our observed $m_{\rm F555W} - m_{\rm F702W}$ color,
as well as the $m_{\rm F25SRF2} - m_{\rm F555W}$ color.
For constant SFR models, the SFR for each of the two epochs is obtained by
matching the model to the $M_{\rm F555W}$ absolute magnitude. For the
instantaneous burst models we tabulate the total mass of stars (an SFR is not
physically meaningful in this case since the burst is represented by a
delta-function).

We point out that the inferred ages are all highly dependent upon the measured
colors, and should therefore be interpreted as {\it upper limits} due to the
likely presence of reddening across most of the field. For example, a decrease
in the intrinsic $m_{\rm F25SRF2} - m_{\rm F555W}$ color by only a further
0.2~magnitudes will yield ages as low as $6 \times 10^7$~yr for the continuous
SFR models, and $\sim 5 \times 10^6$~yr for the instantaneous burst models.
In summary, although the exact ages are uncertain and are model 
dependent, we do find that the knots span a range in age with two of the 
knots being significantly younger ($\lesssim 5 - 10$~Myr) than the other two
($\sim 100$~Myr).

\subsection{Comparison between the Radio and HST Images}

\begin{figure}
\plotone{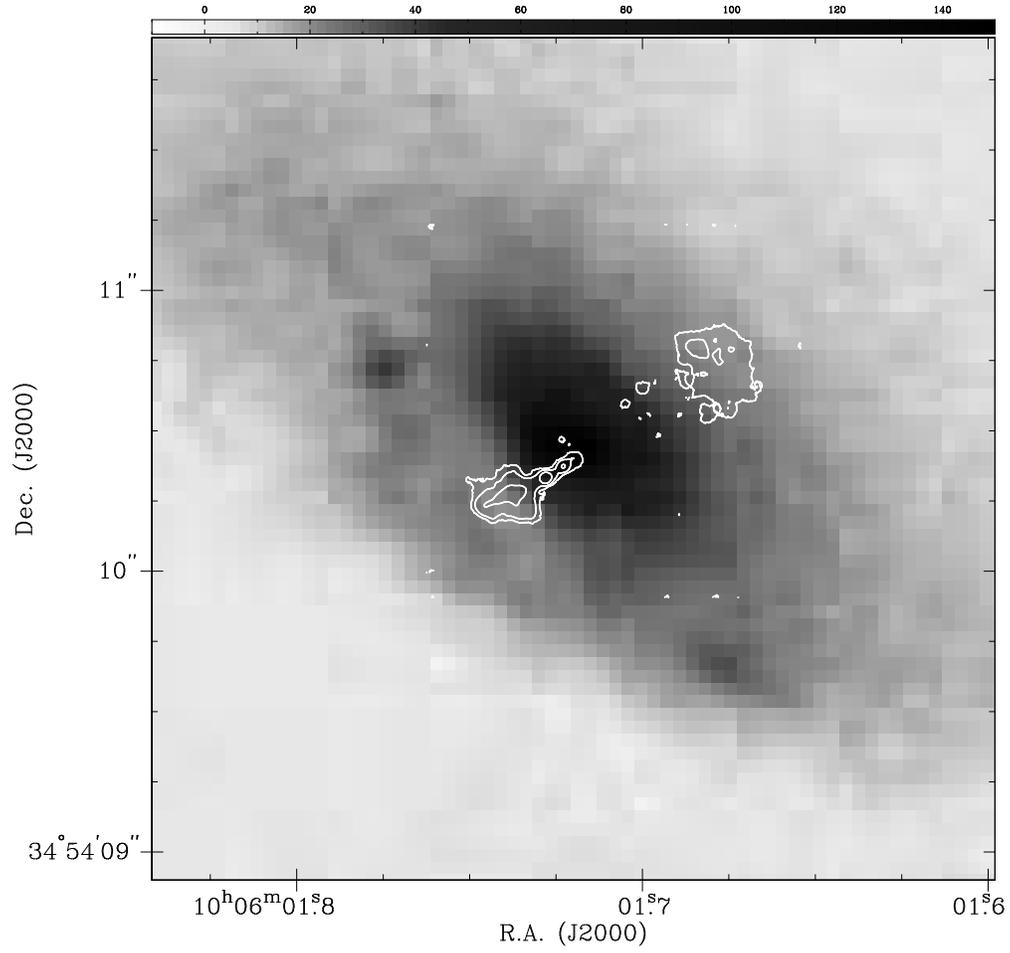}
\caption{3C~236. Overlay of radio image contours (global VLBI 1.66 GHz
image from Schilizzi \etal\ 2000) on the WFPC2 F555W image (grey scale).  } 
\label{figwfpcjetoverlay}
\end{figure}

In Figures~\ref{figwfpcjetoverlay} and~\ref{fig555w702w} we show an overlay 
of the global VLBI 1.66 GHz radio image (Schilizzi \etal\ 2000) superposed 
onto the HST/WFPC2 F555W broad-band and F555W-F702W
color images. The registration between the two images was performed on the the
assumption that the position of the radio core is coincident with the optical
nucleus of the galaxy in the F555W and F702W images (this is slightly different
to the apparent registration in the Schilizzi \etal\  paper where the radio core
is identified with a STIS knot that is to the north-east of the broad-band
nucleus of the galaxy). Furthermore, there is no other strong correspondence
between the kpc-scale radio structure and the blue knots that we detect with
STIS.

\section{DISCUSSION\label{discussion}}

\subsection{The Dust Lane and Inner Disk}

Gas acquired via cannibalism is expected to be smeared into a disk
on a dynamical timescale $t_d \sim 10^8$ yr (e.g., Gunn 1979; 
Tubbs 1980). It will then precess around one of the preferred planes of 
the galaxy, dissipating angular momentum and finally settling into one of
the preferred planes of the galaxy on the precession timescale
$t_p \lae 10^9$ yr (Gunn 1979; Tubbs 1980;
Tohline, Simonson, \& Caldwell 1982; Habe \& Ikeuchi 1985). 
The dissipation of angular momentum will also result
in gas being transported inwards towards the galaxy nucleus (e.g.,
Christodoulou \& Tohline 1993).
In 3C~236 the dust lane seems  asymmetric and filamentary (De Koff 
\etal\ 2000; and Figure~\ref{figdustlane}) and is misaligned with 
the inner disk suggesting the lane is still ``dynamically young". This 
indicates that the host galaxy has recently ($\lae 10^9$ yr) acquired gas 
from a companion. 

3C~236 is in a very poor environment and is considered by Zirbel (1997)
to have no obvious group members within 0.5 Mpc (based on a statistical
correction for background contamination). The donor galaxy for the 
gas could have already been ``eaten". However, we do detect a small
candidate companion galaxy located at 10.1 arcsec (17 kpc) from the 
nucleus along PA 20$\deg$ in our WFPC2 images. 

The mass of dust in the dust lane is estimated by De Koff \etal\ 
(2000) to be $\sim 10^7$ M$_\odot$ based on both HST absorption maps
and IRAS luminosities. The corresponding gas mass is $\sim 10^9$ M$_\odot$
based on the standard gas-to-dust ratio. Thus, the dust lane appears
to be very massive. The large amount of gas in the disk could supply fuel
for a long time and would have allowed the radio galaxy to grow to 
its extremely large size. The relatively sparse environment of the
galaxy may also be an important factor in  producing the large radio 
source. 

The absorption from the dust lane appears to be greater on the south-east
side of the galaxy, consistent with that side of the lane coming out of
the galaxy towards us. This would imply that the north west radio jet
is oriented towards us and the south east jet is oriented away
from us (as also inferred by Schilizzi \etal\ 2000 using De Koff \etal`s
image).  

\subsubsection{Stability of the Alignments}\label{secstab}

A summary of the relevant position
angles are given in Table~\ref{tabangle}.
The orientation of the large scale (4 Mpc) radio source is 122.5$\deg$ 
(Willis \etal\ 1974; Barthel \etal\ 1985). The ``overall" orientation of 
the kpc scale central source is $\simeq 115$ - though there are wiggles 
in the kpc scale jet and departures from this overall orientation -
and on the parsec scale, the orientation of the jet is $\simeq 110\deg$
(Barthel \etal\ 1885; Schilizzi \etal\ 2000). Thus, over the lifetime
of the radio source ($\sim 3\times 10^8$ yr, \S~\ref{secstarradio})
and over a factor of $10^6$ range in size scale, the jets remain
aligned to within about 10$\deg$. 
The large and small scale radio source
are aligned to within a few degrees of perpendicular to  the
``inner" (1 kpc) dust disk which has a projected major axis of 
$\simeq 30\deg$ but are poorly aligned with the perpendicular to the 
larger dust lane which has a projected major axis of $\simeq 55\deg$. 
Thus, the radio source axis is approximately perpendicular to the dust disk
- as found in many radio galaxies (e.g.,
Kotanyi \& Ekers 1979; De Koff \etal\ 1996; Verdoes \etal\ 1999). 

The Bardeen-Petterson effect will cause the black hole to swing its
rotation axis into alignment with the rotation axis of the
disk of gas (on scales of hundreds to thousands of Schwarzschild radii)
which is feeding it; and conversely will keep the spin axis of the inner 
disk aligned with the BH spin (e.g., Bardeen \& Petterson 1975; Rees 
1978). The time scale for alignment of the spin axes is uncertain, 
but for typical parameters expected in the current paradigm
will be in the range $t_{\rm align} \sim 10^6 - 10^7$ yr (Scheuer
\& Feiler 1996; Natarajan \& Pringle 1998). This is much shorter than
the radio galaxy life time and ensures that the jet will be ejected
aligned with the angular momentum vector of the (large scale) 
accretion disk. 

The combination of the long term stability of the jet ejection axis and
the alignment of the jets with the inferred rotation axis of the 
inner kpc-scale dust disk suggests that the orientation of
the inner dust disk has also been stable over the lifetime of the 
radio source.  This is consistent with the inner dust disk being
in a stable resonant orbit - most likely in either an oblate 
(Tohline \etal\ 1982) or triaxial galaxy (Merritt \& de Zeeuw 1983). 
This also implies that the outer misaligned dust lane (which presumably
feeds the disk) settles into the same preferred plane as the disk. 

\subsection{The Star Forming Regions in the Dust Lane}

We note that the 3CR UV snapshot survey  results suggest
that a significant fraction ($\sim 30\%$) of powerful nearby radio galaxies 
show evidence for extended regions of star formation (Allen \etal\ 2000)
-- consistent with early reports of blue colors in  powerful radio galaxies 
(e.g., Smith \& Heckman 1989; McNamara \& O'Connell 1992; McNamara 1995). 
The lack 
of a direct relationship between
the radio source and the the starburst (in contrast to the more powerful
objects at higher redshift) suggests that these are not jet/cocoon-induced
starbursts, but may be ``infall-induced".
This suggests that  (1) infall-induced starbursts are common in nearby 
radio galaxies
and (2) the lifetimes of the starbursts are similar to that of the
radio sources (which is luminosity dependent, but is typically estimated to 
be $10^7-10^8$ yr, e.g., Parma \etal\ 1999). 

The analysis of the star burst colors suggests that the star formation has
been triggered over a range of time scales ranging from perhaps
a $10^{7-8}$ yr for the younger knots to about $10^{8-9}$ yr for the older
knots. The triggering mechanism of the star bursts is not known, but may 
be due to (1) cloud-cloud 
collisions of clouds in adjacent orbits due to differential precession 
in the disk or (2) gas clouds continuing to fall into the disk. 

The star formation in the disk suggests that gas
continues to be transported into the nucleus and possibly fueling
the nuclear activity. However the range of ages of the starbursts
suggests  that the transport of gas into the nucleus may be clumpy and 
sporadic, with different star bursts being triggered by different 
events of infall/transfer of gas. Thus, the existence of non-uniform
transport of gas in the disk may be responsible for both (1) the range
of ages of the star-forming regions, and (2) the apparent episodic nature
of the radio source.

\subsection{The Relationship between the Starbursts and the 
Radio Source}\label{secstarradio}

The large-scale source has a total projected linear size of about
4 Mpc. The projected length of the south eastern side of the source 
is about 24 arcmin (2.4 Mpc). The minimum age of the large scale
source is then
\begin{equation}
t_{min} \simeq 7.8 \times 10^6 \left({v_{lobe} \over c}\right)^{-1} \ \ {\rm yr}
\end{equation}
where $v_{lobe}$ is the advance speed of the lobe. For a canonical
expansion speed of 0.03c (e.g., Scheuer 1985; Alexander \& Leahy 1987),
 the minimum age is $2.6\times 10^8$ yr.
The radio source age is similar to the dynamical time scale in the galaxy
and the age estimates for the older starbursts suggesting that the 
large scale radio galaxy and the initiation of the starbursts are both 
related by a common event, i.e., the infall of gas.

We note that the star bursts appear to be confined to the dust disk
and are not associated with the small scale radio source. Thus,
there does not appear to have been any significant ``jet-induced"
star formation in this source. This is consistent with the hypothesis
that the dense gas
is mostly confined to the plane of the disk and avoids the radio 
source (which seems to be perpendicular to the dust disk). 

Conway (1999) and Conway \& Schilizzi (in preparation) present VLBI 
imaging of the HI absorption towards the
inner radio source. He shows significant HI absorption towards the 
south-east jet/lobe, but none towards the north-west
lobe. This is consistent with the south-east jet being viewed through
the dust lane while the north-west jet is in front of the dust lane. 
Conway (1999) suggests that there is evidence from the kinematics
of the HI for a possible interaction between the south-east  
jet/lobe and the HI gas. This interaction would explain the asymmetry
in the kpc-scale radio source (i.e., the SE side is brighter and shorter than
the NW side). We note that such an interaction, if present, has
not produced any detectable star formation.

\subsection{The Relationship of the Small and Large Radio Sources}

Here we consider the possible relationships between the small (2 kpc)
and large (4.5 Mpc) scale radio sources. The small source could be (1)
brightness enhancements (knots) in the inner part of the large scale jet;
(2) a young source which has resumed activity after a period of dormancy of the 
active nucleus;
(3) the location of the working surfaces of the jet due to infall 
of gas which has smothered the radio jet and temporarily interrupted
the supply of energy to the lobes.

\subsubsection{Bright Inner part of Continuous Jet}

The morphology of the small source argues against the first possibility. 
The small source does not resemble knots in the inner parts of jets.
Instead, we see structures which look like lobes marking the ends of 
the jet. The spectral indices of the small source show a range of values
extending to very steep spectra $\alpha \simeq -1$ (Schilizzi \etal\
2000). These steep spectra are more consistent with those of 
radio lobes than with knots in jets which tend to be $\alpha \simeq -0.65$ 
(e.g., Bridle \& Perley 1984). Schilizzi \etal\ have suggested that
the radio jet extends in a continuous fashion from the core to the lobes 
based on faint emission seen in the low frequency low resolution images. 
 However, (1) an alternate explanation for the faint emission 
is that it is a low surface brightness cocoon or ``bridge" and (2) in a 
continuous
jet scenario there is still an unexplained sudden transition in the jet 
properties at the scale of the kpc source. Thus, for the rest of the 
discussion we consider further only the second and third scenarios. 

\subsubsection{Young Source}\label{secyoung}

If the central 2 kpc source is a young source, then its age is given by
\begin{equation}
t_{min} \simeq 3.2 \times 10^3 \left({v_{lobe} \over c}\right)^{-1} \ \ 
{\rm yr}
\end{equation}
where, for an advance speed of $v_{lobe} = 0.03$c, the minimum age is 
$1.0 \times 10^5$ yr. This is much younger than the youngest of the
star forming regions in the dust lane.  Note that a difference in the
ages is to be expected since the star formation occurs on the kpc 
scale and the jet is fed on the sub-parsec scale. 

\begin{figure}
\plotone{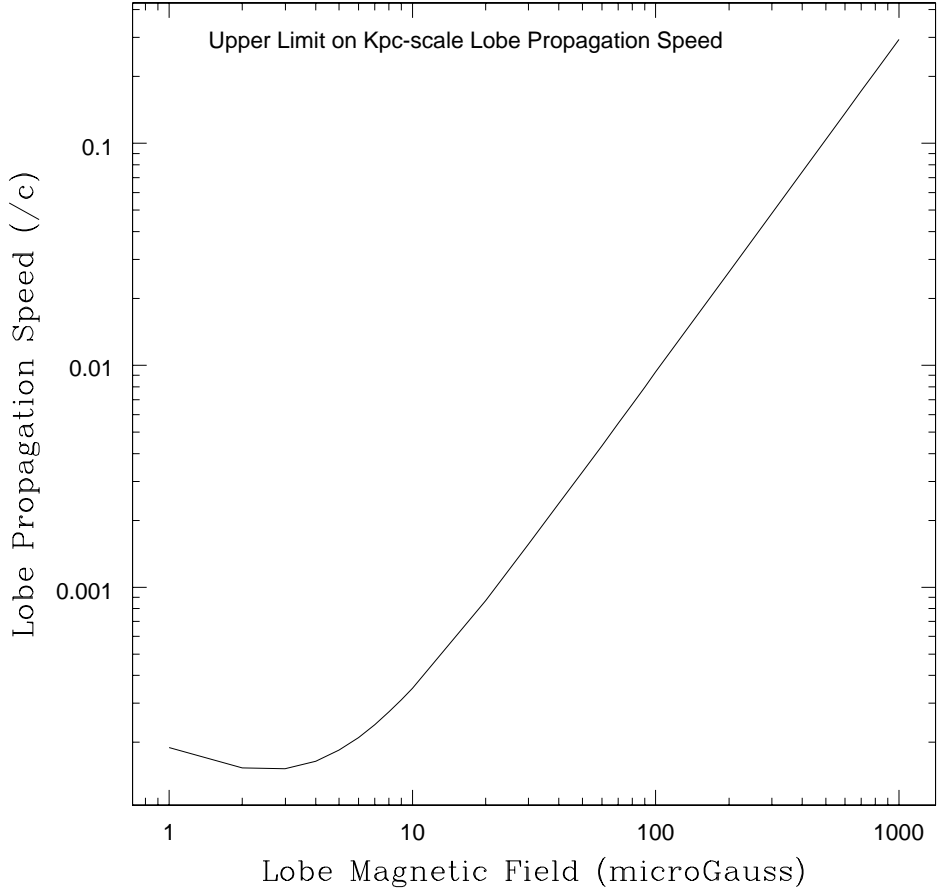}
\caption{3C~236. Plot of the upper limit on the lobe expansion velocity
of the kpc-scale radio source 
as a function of lobe magnetic field, assuming that the source age 
is given by the radiative loss time (eqn.~\ref{eqtrad}). At the
lobe equipartition magnetic field of a few hundred $\mu$G, the lobe
expansion velocity is less than a few percent of c. }
\label{figvelocity}
\end{figure}

We assume that the steep spectra in the small source found by Schilizzi
\etal\ indicate radiative aging of the electrons, rather than an 
intrinsically steep electron spectrum. This implies that the
dynamical age of the source is greater than its radiative age.
For a given magnetic field, this gives an upper limit on the lobe
advance speed $v_{lobe} < l/t_{rad}$ where $l$ is the distance between
the core and hot spot and $t_{rad}$ is the electron radiative lifetime,
due to synchrotron and inverse Compton losses
\begin{equation}\label{eqtrad} 
t_{rad} \simeq 2.6\times10^4 \left({ B^{0.5} \over 
(B^2 + B_R^2) [(1 + z)\nu]^{0.5} }\right) \ \ {\rm yr}
\end{equation}
where $\nu$ is the frequency (in Hz), $B$ is the magnetic field in
Gauss, and $B_R = 4(1 +z)$ is the equivalent magnetic field of
the microwave background (e.g., van der Laan \& Perola 1969). 
We plot the upper limit to the lobe expansion speed as a function
of magnetic field strength adopting a size of 1 kpc and a frequency
of 5 GHz in Figure~\ref{figvelocity}.  At the equipartition 
magnetic field in the two
lobes A and C of a few hundred $\mu$G the particle lifetime is 
$t_{rad} \lae 10^5$ yr  and the upper limit to the expansion 
velocity is a few percent of the speed of light. 

In the young source scenario, the nucleus becomes dormant possibly due
to a lack of fuel to the central engine, the jets cease, 
and the previously ejected jet material traverses the length of
the source in the light travel time (assuming the jet bulk velocity 
is relativistic) shutting off the energy supply to the hot spots (e.g.,
Baum \etal\ 1990; Schoenmakers \etal\ 2000a,b; Kaiser \etal\ 2000). 
We assume that the particles are accelerated primarily in
the hot spots (perhaps by Fermi acceleration in the Mach disk  -
Blandford \& Ostriker 1978 or MHD turbulence - De Young 2000) and that
the acceleration will cease when the jet no longer feeds the hot spot. 
Thus, the particle population will age once the hot spots are no longer
fed. 
If the jet bulk velocity is relativistic as commonly assumed in
powerful classical doubles, then the jets will propagate to the 
hot spots in about $6\times 10^6$ yrs. 
Then, the ages of the youngest electrons in the lobe provides
a constraint on the dormancy period of the nucleus. 
This assumes that the spectral aging estimates are dominated  by the
younger and brighter electron population (see also Jones, Ryu, \&
Engel 1999). This appears to be the case
since the age estimated from spectral aging is significantly less than
the dynamical age based on the likely lobe propagation velocity.

The dormancy period could be longer if (1)  the acceleration continues 
for some timescale after the jet ceases to feed the hotspot, (2) the jet 
propagation to the lobes is significantly slower than light speed,
or (3) the spectral aging estimates are in error (see, e.g., Tribble 1993;
Rudnick, Katz-Stone, \& Anderson 1994).  

The radio spectral aging estimates from Mack \etal\ (1998) are consistent
with ages for the electrons in the hot spots of order  $1 \times 10^7$ yr. 
This would imply a dormancy period for the nucleus of $\gae$ 
$1\times 10^7$ yr. 
If this is the correct model, it would imply that even though the
dust lane is present, the flow of material to the nucleus is not 
continuous;  instead it is clumpy and sporadic on time scales of  
$\sim 10^7$ yr. Alternately, there could be an additional 
factor responsible for the duty cycle of the central engine
which is not directly related to the fuel supply (e.g., some
property of the magnetic field in the accretion disk which
varies on time scales of $\sim 10^7$ yr).

\subsubsection{Interrupted Source}

In the third scenario, the nucleus is active continuously;  however,
dense gas clouds fall into the path of the radio jet and block the
passage of both jets. The energy supply to the lobes is then cut off
until the jets are  able to burrow through the clouds and re-establish
a connection to the lobes. 

This scenario requires that both
jets be cut off symmetrically. 
The existence of HI absorption towards the radio source and the
evidence for possible interaction between the south-east 
jet and the HI gas would support the interrupted source scenario. 
However, there is no evidence for interaction of the north-west lobe 
with ambient gas. In general, it seems like it would be more difficult
to block {\it both\/} jets with infalling gas than it would be to 
simply shut off the fuel supply to the engine (i.e., no infall).  

As in the previous scenario, the particles in the hot spots 
will age after the energy supply is cut off. In this scenario, the time
over which the jets have been blocked, the age of the youngest
electrons in the hotspots, and the age of the small source
should all be similar -- i.e.,   $\sim 10^7$ yr adopting the spectral
aging estimates of  Mack \etal\  (1998). This age seems extremely 
long for such a compact source and would imply the existence of a
cocoon of very steep spectrum emission  - 
which has not been detected (Schilizzi \etal\ 2000). 

Thus, in the interrupted source scenario the two jets have been
blocked for about $\sim 10^7$ yr by dense gas. This requires
(1) the presence of dense gas surrounding the jets, and (2)
the age of the kpc-scale source to be $\gae 10^7$ yr. However, as we 
have argued above, there is no evidence that {\it both\/} jets 
have been blocked.
At present, (1) the very old implied age for the compact radio source
and the lack of very steep spectrum radio emission and (2) the lack of 
evidence for strong confinement of both jets by dense gas are
problems for the scenario that the radio source has been interrupted
by infalling gas.  

\subsection{Non-steady Flow in the Disk?}

The fact that the
dust lane is asymmetric, filamentary, and is misaligned with
the inner disk suggests that is  dynamically young. During this
phase it is expected to be precessing around the preferred plane and
dissipating angular momentum and transporting gas inwards to
the nucleus through processes such as viscosity and cloud-cloud 
collisions (e.g., Shlosman, Begelman, \& Frank 1990). The collisions 
of these clumps and filaments in 
the disk may well trigger the observed star formation. 
The range of estimated ages for the star forming regions 
implies that the star formation is triggered over a range of
time scales and is not based on a single event.
Thus, the existence of the observed clumps and filaments in the dust
lane as well as the range of ages of the star formation regions
suggests that the transport of gas through the disk to the nucleus
is clumpy and non-steady. 

As we have discussed, the properties of the inner 2 kpc radio source
are consistent with a scenario in which the fuel supply to the nucleus
was interrupted for $\sim 10^7$ yr and has now been restored. This
apparent intermittancy in the fuel supply is evidence that the
transport of fuel on the sub-parsec scale is also clumpy and 
non-steady. The outflows in young stellar objects and Herbig-Haro objects
also appear to show evidence for non-steady fueling and repetitive
activity (e.g., Reipurth, Raga \& Heathcote 1992; Bally \& Devine 1994;
Eisloffel \& Mundt 1997).  
Thus, we suggest that the range of ages in the star forming regions
and the repetitive activity in the AGN in 3C~236 are both due to 
non-steady transport of gas in the disk on their respective scales.

We note that a one-to-one correspondence in the age of the young inner
radio source and the youngest starburst in the dust lane is not 
necessarily expected since the dynamical time on the hundreds of 
pc scale where the star formation occurs ($\sim 10^7$ yr) is longer 
than the dynamical 
time scale on the sub-pc scale where the fueling of the AGN occurs. 
However, if the young radio source and the youngest star forming
region are both produced by the same ``disturbance" propagating inwards
through the disk,  then we can use the difference in the ages of 
the youngest starburst ($\sim 10^7$ yr) and the inner radio source 
($\sim 10^5$ yr) to constrain the propagation time.
Thus, in the context of this scenario, the time for the gas to be 
transported in from the hundreds of pc scale to the sub-parsec scale 
is no more than $10^7$ yr which is comparable to the dynamical
time on the hundreds of pc scale. If this is correct, it implies
that nature is able to solve the angular momentum problem on these
scales and that clumps of gas are transported inwards on the dynamical 
time scale. 

\section{SUMMARY AND THE BIG PICTURE}

We present HST WFPC2 V and R band images and STIS/MAMA NUV images
which reveal the presence of four star forming regions in the 
dust lane  in the giant (4 Mpc) radio galaxy 3C~236. 
The dust lane appears to be asymmetric and is misaligned with the 
inner disk suggesting it is still dynamically young. The dust lane
appears to be quite massive ($\sim 10^9$ M$_\odot$) (De Koff \etal\
2000) and clumpy. 

Two of the star forming regions are relatively young with ages of 
$\lesssim 5 - 10$~Myr and metallicities above $\sim 0.2Z_{\sun}$, 
while the other two are older with ages ranging up to a maximum 
of $\sim 100$~Myr.   
$\sim 10^9$ yr. We note that the age of the giant radio source is
likely to be $\sim 10^{8-9}$ yr. 
Thus, the star formation appears to be coeval with the
radio source lifetime. The star formation and the radio source
may both be the common results of infall of gas to the host galaxy. 

We consider two hypotheses for the small kpc-scale source in 3C~236.
(1) It may be a young source ($\sim 10^5$ yr)  which has resumed
activity  after a period of nuclear dormancy of about $10^7$ yr 
(based on the spectral ages of the electrons in the radio lobes). 
(2) It may be a continuously active source in which infalling gas 
has interrupted the jet flow
to the lobes for a period of about $10^7$ yr. 
Both the ``young" and ``interrupted" source scenarios require that 
energy is no longer being supplied to the hot spots. This should be 
tested with high resolution radio observations of the hot spots.  
Currently, the rather old age required for the small source and the
apparent lack of a mechanism for confining the radio source make the
interrupted source hypothesis less appealing. 

The simultaneous existence of the small and large radio source is 
consistent with repetitive activity in the nucleus  on time scales of
$\sim 10^7$ yr. The repetitive nuclear activity may be tied to time 
dependence in the infall of 
fuel on the same time scale. This time scale is similar to that of 
the ages of the younger of the star forming knots. 
The range of ages of the star forming regions implies
repetitive triggering/fueling of the star formation activity in 
the dust lane. Thus,
there is evidence for non-steady activity on both size scales
in which clumps of gas are transported inwards to the nucleus
on the dynamical time scale. 

We suggest that the transport of gas in the disk is non-steady over 
a large range of size scales and that this produces both the multiple 
episodes of star formation in the disk as well as the multiple epochs of radio source activity.


\acknowledgements

We are grateful to Richard Schilizzi for sharing his results in 
advance of publication and Paul Wiita for helpful discussions. 
This work was based on observations made with the NASA/ESA Hubble Space 
Telescope, obtained from the data archive at the Space Telescope Science 
Institute. STScI is operated by the Association of Universities for 
Research in Astronomy, Inc. under the NASA contract NAS 5-26555. 
This work was partially supported by STScI grant GO-08275.01-97A.
This research made use of (1) the NASA/IPAC Extragalactic Database
(NED) which is operated by the Jet Propulsion Laboratory, California
Institute of Technology, under contract with the National Aeronautics and
Space Administration; and (2)  NASA's Astrophysics Data System Abstract
Service.


\clearpage

\begin{deluxetable}{llllr}
\tablewidth{0pt}
\tablecaption{Journal of HST Observations  \label{tabHST}}
\tablehead{
\colhead{Date  } &
\colhead{Program ID  } &
\colhead{Instrument  }  &
\colhead{Filter  } &
\colhead{Time } 
\\
\colhead{ } &
\colhead{ } &
\colhead{ } &
\colhead{  } &
\colhead{Sec  }
}
\startdata
1995-05-07  & 5476 & WFPC2/PC1 & F702W & 4 x 140 \\
1996-06-12  & 6348 & WFPC2/PC1 & F555W & 2 x 300 \\
1999-01-03  & 8275 & STIS/NUV-MAMA & F25SRF2 & 1440  \\
\enddata
\end{deluxetable}

\clearpage

\begin{deluxetable}{lcccccccc}
\tablewidth{0pt}
\tablecaption{\label{tab:fluxes}
	Blue Regions in 3C~236}
\tablehead{
ID	& $\Delta\alpha$ & $\Delta\delta$ & Size			& $m_{\rm F25SRF2} (UV) $	& $m_{\rm F555W} (V) $	& $m_{\rm F702W} (R)$	& $m_{\rm F25SRF2} - $	& $m_{\rm F555W} -$ \\
	& (\arcsec)	 & (\arcsec)	  & (\arcsec)			& (mag)			& (mag)			& (mag)			& $m_{\rm F555W}$	& $m_{\rm F702W}$
}
\startdata
1	& $+0$\farcs8	& $+0$\farcs8	& 0\farcs2			& 25.91$\,\pm$0.07	& 26.74$\,\pm$0.36	& 25.95$\,\pm$0.23	&    $-0.83$\,$\pm$0.43	& \phs$0.79$\,$\pm$0.58	\\
2	& $+0$\farcs6	& $+0$\farcs3	& 0\farcs3			& 23.89$\,\pm$0.03	& 23.76$\,\pm$0.08	& 24.05$\,\pm$0.09	& \phs$0.13$\,$\pm$0.11	&    $-0.29$\,$\pm$0.17	\\
3	& $-0$\farcs1	& $-0$\farcs3	& 0\farcs4$\times$0\farcs2	& 24.04$\,\pm$0.03	& 26.44$\,\pm$0.30	& 26.70$\,\pm$0.34	&    $-1.40$\,$\pm$0.34	&    $-0.26$\,$\pm$0.64	\\
4	& $-0$\farcs4	& $-0$\farcs5	& 0\farcs3$\times$0\farcs2	& 23.91$\,\pm$0.03	& 23.12$\,\pm$0.06	& 22.87$\,\pm$0.05	& \phs$0.79$\,$\pm$0.09	& \phs$0.25$\,$\pm$0.11	\\
nuc	& $0$		& $0$	 	& \nodata			& 23.88$\,\pm$0.03	& 20.93$\,\pm$0.02	& 20.19$\,\pm$0.01	& \phs$2.95$\,$\pm$0.05	& \phs$0.74$\,$\pm$0.04
\enddata
\tablecomments{For each object, the offsets in Right Ascension and Declination
($\Delta\alpha$ and $\Delta\delta$ respectively) are given in arc-seconds
relative to the continuum nucleus of the galaxy. The 1~$\sigma$ uncertainties
are derived from the count-rate statistics.}
\end{deluxetable}

\clearpage

\begin{deluxetable}{ccccc}
\tablewidth{0pt}
\tablecaption{\label{tab:starform}
	Star Formation Parameters for the Blue Knots}
\tablehead{
Model		& \multicolumn{2}{c}{Regions 1 \& 3}	& \multicolumn{2}{c}{Regions 2 \& 4}
}
\startdata
Const. SFR:	& $\log(\tau/{\rm yr})$	& SFR (\Msun/yr)	& $\log(\tau/{\rm yr})$	&  SFR (\Msun/yr)	\\
\tableline
  \Zsun\phm{0.02}	& $7.0 - 7.5$	& $3.1 - 1.7$	& $8.0 - 9.0$	& $2.1 - 0.8$	\\
  0.2\Zsun\phm{0}	& $7.1 - 7.7$	& $2.5 - 1.5$	& $8.0 - 9.0$	& $1.7 - 0.6$	\\
  0.02\Zsun\phm{}  	& $7.3 - 7.8$	& $2.1 - 1.4$	& $8.2 - 9.5$	& $1.0 - 0.4$	\\
\tableline \\
Inst. Burst:	& $\log(\tau/{\rm yr})$	& $\log(M/$\Msun)	& $\log(\tau/{\rm yr})$	& $\log(M/$\Msun)	\\
\tableline
  \Zsun\phm{0.02}	& $6.6 - 7.3$	& $8.5 - 9.3$	& $7.1 - 8.0$	& $8.5 - 9.3$	\\
  0.2\Zsun\phm{0}	& $6.4 - 7.2$	& $8.3 - 9.3$	& $7.3 - 8.8$	& $8.3 - 9.3$	\\
  0.02\Zsun\phm{}	& $6.4 - 6.9$	& $8.8 - 9.3$	& $7.5 - 8.2$	& $8.8 - 9.3$	\\
\enddata
\tablecomments{
The upper and lower limits of the timescale $\tau$ of each model
correspond to the epochs in the model evolution that produce the observed range
of $m_{\rm F25SRF2} - m_{\rm F555W}$ colors. For the Constant SFR models, the
tabulated SFR is that which is required to match the observed integrated magnitude
of the knots. For the Instantaneous Burst models, we instead tabulate the total
mass of stars since in this model all the star formation occurs at one instant in
time. Further details are given in the text.}
\end{deluxetable}

\begin{deluxetable}{lll}
\tablewidth{0pt}
\tablecaption{Position Angles  \label{tabangle}}
\tablehead{
\colhead{Feature  } &
\colhead{Position Angle  } &
\colhead{ Ref  }  
\\
\colhead{ } &
\colhead{Degrees } &
\colhead{ } 
}
\startdata
Kpc-scale  Radio Source & 115 & 1,2 \\
Mpc-scale  Radio Source  & 122 & 1,3 \\
inner dust disk &  30 & 4,5 \\
outer dust lane &  55 & 4,5,6 \\
host galaxy     &  45 & 4 \\
\enddata
\tablerefs{1. Barthel \etal\ (1985). 2. Schilizzi \etal\ (2000). 
3. Willis \etal\ (1974), 4. This paper. 5. De Koff \etal\ (2000).  
6. Martel \etal\ (1999). }
\end{deluxetable}

\clearpage


\end{document}